
%
%
\input harvmac
\def\footsym{*}\def\footsymbol{}\ftno=-2
\def\foot{\ifnum\ftno<\pageno\xdef\footsymbol{}\advance\ftno by1\relax
\ifnum\ftno=\pageno\if*\footsym\def\footsym{$^\dagger$}\else\def\footsym{*}\fi
\else\def\footsym{*}\fi\global\ftno=\pageno\fi
\xdef\footsymbol{\footsym\footsymbol}\footnote{\footsymbol}}
\def\refmark#1{${}^{\refs{#1}}$\ }

\def\hf{\textstyle{1\over2}}
\def\ii{I\bar I}

\def\eg{$e.g.$}

\sequentialequations
\def\sanom{\sigma_{\scriptscriptstyle\rm anom}}
\lref\Othreerefs{For a review and references to the literature
on this model, see V. Novikov,
M. Shifman, A. Vainshtein and V. Zakharov, \sl Phys.~Rep.~\bf116\rm,
103 (1984).  A solution of the model (without a conformal breaking
term) is given by P. Wiegmann,
\sl Phys.~Lett.~\bf152B\rm, 209 (1985).}
\lref\riddle{For a thorough review of the field, see
M. Mattis, \sl Phys.~Rep.~\bf214\rm(3), 159 (1992).}
\lref\SV{Similar-sounding  words can be found in
E. Shuryak and J. Verbaarschot, \sl Phys.~Rev. Lett.~\bf68\rm, 2576
(1992). These authors maintain that the unmodified Khoze-Ringwald
model itself\refmark{\KRone,\KRtwo} bifurcates at an energy where
$\sanom$ is still exponentially suppressed. However, upon close
examination, we are convinced that what they have found is not a
bifurcation at all, but rather an uninteresting coordinate singularity
resulting from their poor choice of $\ii$ collective coordinates.}
\lref\Shuryakold{See also E. Shuryak, \sl Phys.~Lett.~\bf153B\rm, 162 (1985);
\sl Nucl.~Phys.~\bf B302\rm, 559, 574, 599 (1988).}
\lref\DM{N. Dorey and M. Mattis, \sl Phys.~Lett.~\bf B277\rm, 337 (1992).}
\lref\KKR{V. Khoze, J. Kripfganz and A. Ringwald,
CERN preprint CERN-TH.6295/91.}
\lref\KKRII{V. Khoze, J. Kripfganz and A. Ringwald,
CERN preprint CERN-TH.6311/91.}
\lref\MW{E. Mottola and A. Wipf, \sl Phys.~Rev.~\bf D588\rm, 588 (1989).}
\lref\Dorey{N. Dorey, \sl Phys.~Rev.~\bf D46\rm, 4668 (1992).}
\lref\BalBraun{I. Balitskii and V. Braun,
\sl Nucl.~Phys.~\bf B380\rm, 51 (1992).}
\lref\Trieste{M. Mattis, in the
 Proceedings of the ICTP Summer School in High Energy Physics and
Cosmology, Trieste 1991, World Scientific (to appear).}
\lref\DHN{R. Dashen, B. Hasslacher and A. Neveu, \sl Phys.~Rev.~\bf D10\rm,
4138 (1974).}
\lref\Voloshininit{M. Voloshin, \sl Nucl.~Phys.~\bf B359\rm, 301 (1991).}
\lref\Yaffe{L. Yaffe, in  \it
Baryon Number Violation at the SSC? Proceedings of the Santa Fe
Workshop\rm, eds. M. Mattis and E. Mottola, Singapore, World
Scientific, 1990.  }
\lref\Shuryak{E. Shuryak, \sl Nucl.~Phys.~\bf B203\rm, 93, 116, 140 (1982).}
\lref\tHooft{G. 't Hooft, \sl Phys. Rev. \bf D14\rm, 3432 (1976);
\bf D18\rm, 2199(E) (1978).}
\lref\Ringwaldbook{A. Ringwald, in Ref.~\book.}
\lref\BPST{A. Belavin, A. Polyakov, A. Schwartz and Y. Tyupkin, \sl
Phys.~Lett.~\bf B59\rm, 85 (1975).}
\lref\cutting{R.~Cutkosky, \sl J.~Math.~Phys. \bf 1\rm,
429 (1960). The clearest
introduction to cutting rules is G.~'t Hooft and T.~Veltman, \sl
Diagrammar\rm, pp. 210-217, in D.~Speiser \etal, eds., \sl Particle
Interations at Very High Energies\rm, Part B (1974).}
\lref\Goldberg{H. Goldberg, \sl Phys.~Lett.~\bf B246\rm,
445 (1990), and contribution
to Ref.~\book.}
\lref\AMgauge{P. Arnold and M. Mattis, \sl Mod.~Phys.~Lett.~\bf A6\rm, 2059
(1991).}
\lref\KT{S. Khlebnikov and P. Tinyakov,
\sl Phys.~Lett.~\bf269B\rm, 149 (1991).}
\lref\KRTOthree{S. Khlebnikov, V. Rubakov
and P. Tinyakov, \sl Nucl.~Phys.~\bf B347\rm, 783 (1990).}
\lref\initial{M. Mattis, L. McLerran and L. Yaffe, \sl Phys.~Rev.~\bf D45\rm,
4294 (1992); S.~Khlebnikov, UCLA-91-TEP-38 (1991); V.~Rubakov and P. Tinyakov,
\sl Phys.~Lett.~\bf B279\rm, 165 (1992); P. Tinyakov,
\sl Phys.~Lett.~\bf B284\rm, 410 (1992); A. Mueller, CU-TP-572 (1992).}
\lref\Mueller{A. Mueller, \sl Nucl. Phys. \bf B348\rm, 310 (1991).}
\lref\JacReb{R. Jackiw and C. Rebbi, \sl Phys.~Rev.~\bf D14\rm, 517 (1976).}
\lref\Aoyama{H. Aoyama and H. Kikuchi, \sl Phys.~Lett.~\bf B247\rm, 75 (1990);
\sl Phys.~Rev.~\bf D43\rm, 1999 (1991).}
\lref\Ringrev{For  recent reviews focusing on the use of the valley method in
high-energy scattering, see A. Ringwald, Proceedings of the 1991 PASCOS
meeting,
and V. Khoze, Proceedings of the INFN Eloisatron Project, 17th Workshop ``QCD
at 200 TeV,'' June 1991.}
\lref\Verbaarschot{J. Verbaarschot, \sl Nucl.~Phys.~\bf B362\rm, 33 (1991).}
\lref\Cornwall{J. Cornwall, \sl Phys.~Lett.~\bf B243\rm,
271 (1990), and contribution
to Ref.~\book.}
\lref\Zakharov{V. Zakharov, \sl Nucl.~Phys.~\bf B353\rm, 683 (1991) and
\sl Nucl.~Phys.~\bf B377\rm, 501 (1992).}
\lref\Muellernfac{T. Lee and A. Mueller, Columbia preprint CU-TP-523 (1991).}
\lref\Ringwald{A. Ringwald, \sl Nucl. Phys. \bf B330\rm, 1 (1990).}
\lref\Espinosa{O. Espinosa,  \sl Nucl. Phys. \bf B343\rm, 310 (1990).}
\lref\MVV{L. McLerran, A. Vainshtein and M. Voloshin,
\sl Phys. Rev. \bf D42\rm, 171 (1990).}
\lref\MVVII{L. McLerran, A. Vainshtein and M. Voloshin,
\sl Phys. Rev. \bf D42\rm, 180 (1990).}
\lref\KRT{S. Khlebnikov, V. Rubakov and P. Tinyakov,
\sl Nucl. Phys. \bf B350\rm, 441 (1991).}
\lref\new{P. Arnold and M. Mattis, \sl
Phys. Rev. Lett. \bf66\rm, 13 (1991).}
\lref\KRone{V.~V.~Khoze and A. Ringwald, \sl Nucl. Phys. \bf B355 \rm
351 (1991).}
\lref\KRtwo{V.~V.~Khoze and A. Ringwald, \sl Phys. Lett. \bf B259 \rm
106 (1991).}
\lref\danger{J. Bossart and C. Wiesendanger, \sl Phys.~Rev.~\bf D46\rm,
1820 (1992).}
\lref\KhozeIII{V.~V.~Khoze and A. Ringwald,  CERN preprint
CERN-TH-6082/91 (1991).}
\lref\Balitsky{I. Balitsky and A. Yung, \sl Phys.
Lett. \bf168B\rm, 113 (1986).}
\lref\Yung{A. Yung, \sl Nucl. Phys. \bf B297\rm, 47 (1988).}
\lref\Novikov{M. Shifman, A. Vainshtein and V. Zakharov, \sl Nucl.~Phys.~\bf
B165\rm, 45 (1980); V. Novikov, M. Shifman, A. Vainshtein and V. Zakharov,
\sl Sov.~Phys.~Usp. \bf25\rm, 195 (1982).}
\lref\KRTII{S. Khlebnikov, V. Rubakov and
P. Tinyakov, Institute for Nuclear
Research Report INR-TH-1001-90 (1990).}
\lref\YungII{A.~Yung, Leningrad preprint LNPI-1617 (1990).}
\lref\Diakonov{D. Diakonov and V.~Petrov, in the Proceedings of the 26th
Winter School of the Leningrad Institute
of Nuclear Physics, Leningrad 1991
(in Russian).}
\lref\MuellerII{A. Mueller, \sl Nucl. Phys. \bf B353\rm, 44 (1991).}
\lref\Shuryak{E. Shuryak, \sl Nucl.~Phys.~\bf B203\rm, 93, 116, 140 (1982).}
\lref\Banks{T.~Banks, G.~Farrar, M.~Dine, D.~Karabali, B.~Sakita, \sl
Nucl. Phys. \bf B347\rm, 581 (1990), and in Ref.~\book.}
\lref\Li{X.~Li, L.~McLerran, M.~Voloshin and
R.~Wang, \sl Phys.~Rev.~\bf D44\rm, 2899 (1991).}
\lref\LY{H. Levine and L. Yaffe, \sl Phys. Rev. \bf D19\rm,
1225 (1979).}
\lref\ZakNP{V. Zakharov, \sl Nucl.~Phys.~\bf B353\rm, 683 (1991).}
\lref\Porrati{M. Porrati, \sl Nucl. Phys. \bf B347\rm, 371 (1990).}
\lref\AM{P. Arnold and M. Mattis, \sl Phys. Rev. \bf D42\rm, 1738 (1990).}
\lref\Voloshin{M. Voloshin, TPI preprint TPI-MINN-91/5-T (1991).}
\lref\Muellernew{A. Mueller, \sl Nucl.~Phys.~\bf B364\rm, 109 (1991).}
\lref\Bern{Z. Bern, private communication.}
\lref\Ramond{See for example Sec. III.3 of
P. Ramond, \sl Field Theory: A Modern Primer\rm,
Benjamin Cummings, 1981.}
\lref\BCCL{L. Brown, R. Carlitz, D. Creamer and C. Lee, \sl
Phys. Rev. \bf D17\rm, 1583 (1978).}
\lref\Manton{N. Manton, \sl Phys.~Rev.~\bf D28\rm, 2019 (1983); F. Klinkhamer
and N. Manton, \sl Phys.~Rev. \bf D30\rm, 2212 (1984).}
\lref\Coleman{S. Coleman, \sl The Uses of Instantons\rm, in the Proceedings
of the 1977 Erice Summer School, ed.~A. Zichichi; reprinted in S. Coleman,
\sl Aspects of Symmetry\rm, Cambridge Univ.~Press 1985.}
\lref\KRTperiodic{S. Khlebnikov, V. Rubakov and P. Tinyakov, DESY preprint
DESY-91-033 (1991).  Here it
is shown that periodic instantons have nothing to do with the problem of
high-energy $B$ violation.}
\lref\Farrar{G. Farrar and R. Meng, \sl
Phys.~Rev.~Lett.~\bf65\rm, 3377 (1990).}
\lref\GPY{D. Gross, R. Pisarski and L. Yaffe, \sl Rev.~Mod.~Phys.~\bf53\rm,
43 (1981).}
\lref\Kuzmin{V. Kuzmin, V. Rubakov and M.
Shaposhnikov, \sl Phys.~Lett.~\bf155B\rm,
36 (1985).}
\lref\ArnMcL{P. Arnold and L. McLerran, \sl Phys.~Rev.~\bf D36\rm,
581 (1987)}
\lref\ArnMcLII{P. Arnold and L. McLerran, \sl Phys.~Rev.~\bf
 D37\rm, 1020 (1988).}
\lref\ArnoldEmil{For recent reviews of the related
 problem of anomalous $B$ violation
at high temperatures, see M. Shaposhnikov, Proceedings of the ICTP
Summer School in High Energy Physics and Cosmology, Trieste, 1991; P.
Arnold, Proceedings of the 1990 TASI Summer School; E. Mottola,
Proceedings of the 1989 TASI Summer School.}
\lref\Affleck{I. Affleck, \sl Nucl.~Phys.~\bf B191\rm, 445 (1981).}
\lref\Maggiore{M. Maggiore and M. Shifman, \sl Nucl.~Phys.~\bf B371\rm, 177
(1992) and  \sl Phys.~Rev.~\bf D46\rm, 3550 (1992).}
\lref\Shifman{M. Maggiore and M. Shifman,
TPI preprints TPI-MINN-91/24-T and 91/27-T (1991).}
\lref\peace{P. Arnold and M. Mattis, \sl Phys.~Rev.~\bf D44\rm, 3650 (1991).}
\lref\Voloshinbubble{M. Voloshin, \sl Nucl.~Phys.~\bf B363\rm, 425 (1991).}
\lref\ABJ{S. Adler, \sl Phys.~Rev.~\bf177\rm, 2426 (1969); J. Bell and R.
Jackiw, \sl Nuovo Cim.~\bf60A\rm, 47 (1969).}
\lref\SLC{G. Abrams, \etal, \sl Phys.~Rev.~Lett.~\bf63\rm, 2173 (1989).}
\lref\DiakonovII{D. Diakonov and M. Polyakov,
Leningrad LNPI preprint 1737 (1991).}
\Title{LA-UR-93-811}{\vbox{\centerline{Valley Bifurcation in an $O(3)$
$\sigma$ Model:}\centerline{Implications
 for High-Energy Baryon Number Violation}}}

\centerline{\authorfont Jeffrey M. Grandy and
Michael P. Mattis\footnote{$^\dagger$}{mattis@pion.lanl.gov}}
\bigskip
\centerline{\sl Theoretical Division T-8, Los Alamos National Laboratory}
\centerline{\sl Los Alamos, NM 87545}

\vskip .3in
The valley method for computing the total high-energy anomalous cross
section $\sanom$ is the extension of the optical theorem to the case of
instanton-antiinstanton backgrounds.  As a toy model for baryon number
violation in Electroweak theory, we consider a version of the $O(3)$
$\sigma$ model in which the conformal invariance is broken perturbatively.
We show that at a critical energy the saddle-point values of the instanton
size and instanton-antiinstanton separation bifurcate into complex
conjugate pairs.  This nonanalytic behavior signals the breakdown of the
valley method at an energy where $\sanom$ is still exponentially suppressed.
\vskip .3in
\Date{February 1993} 
\vfil\break
\newsec{Introduction}

Despite intense theoretical effort,\refmark\riddle the riddle of
high-energy baryon number violation remains unsolved nearly four years
after the original calculations of Ringwald\refmark\Ringwald and
Espinosa.\refmark\Espinosa The phenomenon is more or less the same in
2-dimensional systems such as the abelian Higgs model, or the $O(3)$
$\sigma$ model which we focus on here, as it is in 4 dimensions in
Weinberg-Salam.  In each case, one calculates for the inclusive
anomalous $2\rightarrow\,$many cross section:\refmark{\AM,\KRT,\Yaffe}
\def\fhg{F_{hg}}
\def\scl{S_{cl}}
\eqn\sanomdef{\sanom\ \sim\ \exp\big(2\scl\cdot\fhg(E/E_s)\, \big)\ ,}
neglecting sub-exponential effects, which vary much more slowly with
energy.
Here $\scl$ is the action of a single (small) instanton, and $\fhg,$ the
so-called ``holy grail function,'' is a rising function of energy
measured in units of a characteristic scale $E_s$ of order the sphaleron
mass. $\fhg$ has the general form
\eqn\fhgdef{\fhg(E/E_s)\ =\ -1+c_1(E/E_s)^{k_1}+
\big(c_2+l_2\log(E/E_s)\big)\cdot
(E/E_s)^{k_2} +\cdots\, .}
Only the constants $c_i$, $l_i$ and $k_i$
are model-dependent. In Weinberg-Salam, $k_1=4/3$ with subsequent
$k_i$ increasing by 2/3, while in the $O(3)$ $\sigma$ model $k_1=1$
with subsequent $k_i$ increasing by unity.\refmark\KRT

The riddle in all these models is: Does the holy
grail function rise close enough to zero that the exponential suppression
is lost and $\sanom$ becomes observable? A closely related question is:
What is the mechanism that keeps $\fhg$ from becoming positive, yielding
an exponentially large $\sanom$ in flagrant violation of the unitarity
bounds of quantum field theory?

A useful \it approximate\rm\foot{\rm We comment on multi-instanton and
initial-state corrections, ignored in our treatment, at the end.}
\rm tool  for examining these issues is the valley
method of Balitsky and Yung,\refmark{\Balitsky,\Yung,\Shuryakold}
 adapted to high-energy scattering by Khoze
and Ringwald.\refmark{\KRone,\KRtwo}
  In the Khoze-Ringwald approach, $\sanom$ is extracted via the optical
theorem as the imaginary part of a $non$anomalous forward $2\rightarrow2$
amplitude in which the intermediate state contains a distorted
instanton-antiinstanton ($\ii$) pair. The set of such $\ii$ configurations
satisfying the appropriately
constrained Euler-Lagrange equation (``valley
equation''\refmark{\Balitsky,\Yung}) is known
as the valley. In both the $O(3)$ $\sigma$ model\refmark{\danger,\Dorey} and in
Weinberg-Salam,\refmark{\KRone,\KRtwo}
 $\fhg$ from Eq.~\sanomdef\ is then approximated as the sum of three terms:
\def\sconf{S_{\rm conf}}
\eqn\newfhg{\sanom\ \sim\ \hbox{Im}\,\int\,dR\,d\rho\,
\exp\left\{E R-\sconf( R/\rho)
-\scl\cdot2\rho^2\mu^2\right\}\  ,}
where the  collective coordinate integrations over $\rho$ (the
(anti)instanton size) and
$R$ (the $\ii$ separation)
 are to be evaluated  in saddle-point approximation. Taking the
imaginary part in Eq.~\newfhg\ strips off the factor of $i$ that
enters by analytic continuation
from the ``wrong-sign'' Gaussian integral $\int dx\, e^{+ax^2}$
implicit in the small-fluctuations determinant about the saddle
point.

The three terms in Eq.~\newfhg\ have the following interpretation.
The first term on the right-hand side is the proper Euclidean continuation
of the phase factor due to pumping energy $E$ into the system through
the initial-state quanta. The second and third terms represent a
splitting-up of the valley action into a classically
conformally invariant piece
$\sconf$ depending only on the dimensionless ratio $R/\rho,$ and an
ad-hoc
conformal breaking term $2\rho^2\mu^2,$ where
 $\mu$ is a characteristic mass of order $g^2E_s$
(\eg, $\mu=\hf M_W$ in Weinberg-Salam), and
the factor of 2 reflects
the identical contributions of the $I$ and the $\bar I.$  For example,
in Weinberg-Salam, $\sconf$ comes from the pure Yang-Mills part of the
action, whereas the last term (which is only justified when $\rho\ll
M_W^{-1}$, see Ref.~\Affleck) crudely models the effect of the Higgs.

The upshot of the Khoze-Ringwald approach,
in both the $O(3)$ $\sigma$ model\refmark\Dorey
and in Weinberg-Salam,\refmark\KRtwo is the following. In the one-instanton
sector of the theory, $\fhg$ rises monotonically with energy, starting
at $-1$ as per Eq.~\fhgdef, and hitting zero at some critical energy
$E_{KR}$ of order $E_s.$  For $E\ge E_{KR}$ the Khoze-Ringwald method breaks
down, and extra physics is needed, but this is a moot point. For, the
Khoze-Ringwald
scenario has not only predicted that $\sanom$ loses its exponential
suppression at some finite energy potentially accessed by experiment,
it has also provided a putative mechanism for keeping $\fhg$ from
becoming positive, thus ensuring unitarity.

In this Letter, we improve on the Khoze-Ringwald approach in a
definite way. Specifically, for the first time in any model, we
promote the ad-hoc conformal breaking term $2\rho^2\mu^2$ in
Eq.~\newfhg\ to a bona fide term in a Lagrangian.  In the particular
version of the $O(3)$ $\sigma$ model that we examine, we then find an
altogether different behavior than the Khoze-Ringwald scenario,
namely, a \it bifurcation \rm in the valley at an energy at which
$\sanom$ is still exponentially suppressed.\refmark\SV By a
bifurcation, we mean that the saddle-point values of $\rho$ and $R$
leave the real axis as complex-conjugate pairs, at which point the
optical-theorem justification of the valley
method\refmark{\peace,\BalBraun} is apparently lost.

This bifurcation scenario was first outlined in Sec.~4 of Ref.~\DM.
The present work fleshes it out in the context of a specific
Lagrangian model.  Whether a similar bifurcation holds for other
versions of the $O(3)$ $\sigma$ model in which conformal breaking is
handled differently, or indeed in Weinberg-Salam when the effect of
the Higgs sector on the valley is treated correctly, is anyone's
guess. But it is at least \it plausible \rm that the phenomenon we
exhibit here turns out to be more general, and furthermore, that it
prevents $\sanom$ from ever becoming observable at sphaleron energies.

\newsec{Motivating the model}

The classically
conformally invariant $O(3)$ $\sigma$ model is defined by the Euclidean
action\refmark\Othreerefs
\eqn\sconfdef{\eqalign{\sconf\ &=\ {1\over2g^2}
\int d^2x\,\partial_\mu\hat{\bf n}\cdot
\partial_\mu\hat{\bf n}\cr&=\ {1\over2g^2}\int{dz\,d\bar z}{1\over
{(1+w\bar w)}^2}\left({\partial w\over\partial z}
{\partial \bar w\over\partial \bar z}+{\partial w\over\partial \bar z}
{\partial \bar w\over\partial z}\right)\ }}
where $\hat{\bf n}$ lives on the 2-sphere.
In the second equality we have passed to the complex representation
\eqn\comprep{w={\hat n_1+i\hat n_2\over1-\hat n_3}\ ,\quad z=x_1+ix_0\ .}
While the $O(3)$ symmetry is more obscure in this representation, what becomes
manifest is conformal invariance. Specifically,
$\sconf$ is invariant under the 1-to-1 conformal mappings
\eqn\onetoone{z\rightarrow f(z)={az+b\over cz+d}\ ,\quad w(z)\rightarrow
w\big(f(z)\big)\ .}

The $I$'s ($\bar I$'s) in this model have the simple (anti)analytic form
\def\bI{{\bar I}}
\eqn\instdef{w_I\ =\ {\rho_Ie^{i\theta_I}\over z-z_I}+c_I\ ,\quad
w_\bI\ =\ {\rho_\bI e^{i\theta_\bI}\over \bar z-\bar z_\bI}+c_\bI}
and action $4\pi/g^2.$ Here $\rho_I$ ($\rho_\bI$) and $z_I$ ($z_\bI$)
are the (anti)instanton's scale size and location, respectively, while
the phases $\theta_I$ ($\theta_\bI$) and asymptotic constants $c_I$
($c_\bI$) fill out the $SL(2,C)$ manifold of collective coordinates.
For our purposes, we need also the $I\bar I$ valley. In the notation
of Ref.~\danger, it is given by the concentric configuration
\eqn\concval{w_V(z,\bar z)\ =\ {is\over
u-u^{-1}}z^{-1}+{is^{-1}\over u-u^{-1}}\bar z\ ,}
followed by any of the transformations $z\rightarrow f(z)$ given in
Eq.~\onetoone\ (see Refs.~\danger-\Dorey\ for details).  It then
appears that $w_V$ depends on an unmanageably large number of complex
parameters.  Fortunately, most of them are redundant.  To be precise,
by rotating the phases of $z$ and $\bar z$, and translating and
factoring a complex phase from $w_V$, we can actually express $w_V$ in
terms of just three real collective coordinates $(\rho_I,\rho_\bI,R)$:
\eqn\newval{w_V(z,\bar z)\ =\ {\rho_I\over R}\cdot{z-R/2\over z+R/2}+
{\rho_\bI\over R}\cdot{\bar z+R/2\over \bar z-R/2}\ ,}
where the new parameter $R$ measures the $I\bar I$ separation. Such
phase redefinitions of $w_V$ are permissible provided that all
expressions of interest (\eg, Eq.~\sconfdef) depend only on real
products such as $w\bar w.$

So far as the $\sconf$ contribution to $\fhg$ is concerned (albeit not
the other two terms in Eq.~\newfhg), the
set $(\rho_I,\rho_\bI,R)$ is redundant still. In fact,
the valley action is\refmark\danger
\eqn\valaction{\sconf(w_V)\ =\ {8\pi\over g^2}
\left[{u^4+1\over{(u^2+1)}^2}-{2u^4\log u^4\over(u^2-1){(u^2+1)}^3}\right]}
where the single
valley parameter $u$ introduced in Eq.~\concval\ is now reexpressed
as\refmark\Dorey
\eqn\udef{u\ =\
\left({R^2+2\rho_I\rho_\bI+R\sqrt{
R^2+2\rho_I\rho_\bI}\over2\rho_I\rho_\bI}\right)^{1/2}
\ .}
$\sconf(w_V)$  interpolates smoothly between the far-separated regime at the
$u\rightarrow\infty$ end of the valley,
\eqn\vallimita{\sconf(w_V)\ \rightarrow\ {8\pi\over
g^2}\ \equiv\  \sconf(w_I)+\sconf(w_\bI)
\quad\hbox{when}\quad
R\ \gg\ \rho_I,\rho_\bI\ ,}
and the perturbative vacuum as $u\rightarrow1$:
\eqn\vallimitb{\sconf(w_V)\ \rightarrow\ 0 \quad
\hbox{when}\quad R\ \ll\ \rho_I,\rho_\bI\ .}
Rather than vanishing as one would expect, $w_V$ as given in
Eq.~\newval\ actually blows up in this latter limit. However, for any
configuration, vanishing and blowing up are really the same thing in
this model, since $w\rightarrow w^{-1}$ (equivalently $\hat
n_2\rightarrow-\hat n_2$, $\hat n_3\rightarrow-\hat n_3$) is a
specific instance of the $O(3)$ symmetry of Eq.~\sconfdef.

\def\scsb{S_{\rm csb}}
In order to serve as a plausible toy model for Electroweak theory,
$\sconf$ needs to be supplemented by an explicit conformal symmetry
breaking term $\scsb.$\refmark\MW The presence of $\scsb$ will modify
the valley equation, and consequently the valley itself (as well as
the $I$'s and $\bar I$'s). Unfortunately, solving for the conformally
broken valley in any field theory is a formidable numerical task.
Recall that for conformally invariant field theories, the concentric
valleys are obtained by a series of mathematical tricks that map the
problem onto a solvable quantum mechanical
model.\refmark{\Yung,\danger} These tricks become invalid when the
conformal symmetry is broken.

To simplify
 our task, we limit ourselves herein to first order perturbation theory.
In other words, we will simply  plug the known conformally invariant
valley \concval-\newval\ into
$\scsb.$ In order to trust this approximation,
 we will verify \it a posteriori \rm
that
\eqn\fopt{\scsb[w_V]\ \ll\ \sconf[w_V]}
throughout the energy range of interest. A comparable
first-order perturbation theory scheme has been tacitly assumed in previous
valley-method calculations.

What to use for $\scsb$? Mottola and Wipf have used\refmark\MW
\eqn\mwterm{\scsb^{\rm MW}\ =\ {4\pi\mu^2\over g^2}\cdot
\int dzd\bar z\,{1\over1+w\bar w}}
in their study of sphaleron physics in this model.  However,
$\scsb^{\rm MW}$ is unsuitable for our purposes, because
it diverges on the instanton \instdef.
Alternatively, Khlebnikov, Rubakov and Tinyakov have used\refmark\KRTOthree
\eqn\krtterm{\scsb^{\rm KRT}\ =\ {4\pi\mu^2\over g^2}\cdot\int dzd\bar z
\,\left({w\bar w\over1+w\bar w}\right)^2}
which is finite on the subset of instantons \instdef\ for which the asymptotic
constant $c_I^{}$ is zero. However, $\scsb^{\rm KRT}$ still
diverges on the valley \newval. The reason is that
\eqn\valbc{w_V(z,\bar z)\ \longrightarrow\ -{\rho_I^{}+\rho_\bI\over R}\quad
\hbox{as}\quad |z|\rightarrow\infty}
and this unavoidable asymptotic constant gives rise to an infrared divergence.

We are led to concoct a term with gradients that kill this asymptotic
constant. A natural set of such terms are powers of the kinetic energy density:
\eqn\scsbdef{\scsb\ =\ {4\pi\over g^2}\cdot
\sum_{n=1}^\infty\,f_n\,\mu^{2-2n}\,
\int{dz\,d\bar z}{\left[{1\over{(1+w\bar w)}^2}
\left({\partial w\over\partial z}
{\partial \bar w\over\partial \bar z}+{\partial w\over\partial \bar z}
{\partial \bar w\over\partial z}\right)\right]}^n\ .}
Whereas the mass scale $\mu$ is inserted for dimensional reasons,
the dimensionless constants $f_n$ can be chosen in any convenient manner.
We will restrict this choice by demanding that at low energies,
$\scsb(w_V)$
reduces to the Khoze-Ringwald form of the conformal breaking
term shown in Eq.~\newfhg. Equation \scsbdef\
is actually more symmetric than either $\scsb^{\rm MW}$ or $\scsb^{\rm KRT}$
as it preserves the $O(3)$ invariance. The fact that it is
 nonrenormalizable does not bother us, as we are focusing
exclusively on semiclassical physics.

For guidance in selecting the constants
$f_n$ intelligently, we insert the instanton $w_I=\rho/z$ and calculate
\eqn\instinsert{\scsb(w_I=\rho/z)\ =\ {4\pi\over g^2}\cdot
\sum_{n=1}^\infty\,{\pi f_n\over2n-1}\cdot(\mu\rho)^{2-2n}\ .}
The choice
\eqn\fnchoice{f_n\ =\ -{2n-1\over\pi}\cdot(-\lambda/g^2)^{-n}}
leads to the geometric series
\eqn\scsbinst{\scsb(w_I=\rho/z)\ =\ {4\pi\over g^2}\cdot
{\mu^2\rho^2\over1+(\lambda/g^2)\mu^2\rho^2}\ .}
Of course, what we want is $\scsb(w_V)$, not $\scsb(w_I)$. But
Eq.~\scsbinst\ is suggestive. For, in the low-energy limit, the $I$
and $\bar I$ are well separated, and we therefore expect that
$\scsb(w_V)\rightarrow2\scsb(w_I)$, as happens for $\sconf$ ($cf.$
Eq.~\vallimita). Furthermore, in this limit, the saddle-point value of
$\mu^2\rho^2\rightarrow0,$\refmark{\KRTOthree,\KRT} so that the
denominator in Eq.~\scsbinst\ approaches unity.  Therefore,
$\scsb(w_V)$ specified by Eqs.~\scsbdef\ and \fnchoice\ pleasingly
reduces at low energies to the Khoze-Ringwald form of the conformal
breaking term shown in Eq.~\newfhg\ (as we explicitly verify below).
The reader can check that imposing this low-energy limit forces us to
take an infinite number of powers of the kinetic energy density, so
that the choice \fnchoice\ is in a sense a minimal construction.  At
the same time---and this is the new feature of our
calculation---$\scsb$ provides a well-defined Lagrangian prescription
for extrapolating to higher energies.

 The remaining free parameter $\lambda/g^2$ in
Eqs.~\fnchoice-\scsbinst\ determines the relative strengths of the
$\mu^2\rho^2$ and $\mu^4\rho^4$ contributions to $\scsb.$ In this
respect (and motivating our notation), $\lambda/g^2$ is roughly
analogous to the ratio $\lambda^{}_{\rm higgs}/g_W^2\sim M_{\rm
higgs}^2/M_W^2$ in Electroweak theory,\foot{See Eqs.~(3.27) and (3.4)
in Ref.~\Espinosa} where $\lambda_{\rm higgs}$ is the Higgs
self-coupling and $g^{}_W$ is the $SU(2)_W$ gauge coupling.  Our
freedom to tune $\lambda/g^2$ will turn out to be important in
ensuring that the perturbation theory criterion \fopt\ is met.

\newsec{Results and Discussion}

To recapitulate, our model is defined by
\eqn\ourmodel{2\scl\cdot\fhg\ =\ ER-\sconf(w_V)-\scsb(w_V)}
where $w_V$ is given in Eq.~\newval, $\sconf$ is given in
Eqs.~\valaction-\udef, and $\scsb$ is given in Eqs.~\scsbdef\ and
\fnchoice.

 The values of the $\ii$ collective coordinates $\rho_I,
\rho_\bI$ and $R$ used in Eqs.~\newval\ and \ourmodel\ are to be determined
self-consistently from Eq.~\ourmodel\ by  saddle-point methods.
To simplify this task, we introduce the rescaled dimensionless variables
\eqn\newvars{\eqalign{\theta\ &=\ \rho/R\cr \zeta\ &=\ (\mu R)^{-2}\cr
\epsilon\ &=\ g^2E/4\pi\mu}}
where, as in all previous work on the valley method, we have anticipated that
by symmetry
\eqn\symrhos{\rho_I=\rho_\bI\equiv\rho}
at the saddle point. Equation \ourmodel\ then becomes
\def\tsconf{\tilde S_{\rm conf}}
\def\tscsb{\tilde S_{\rm csb}}
\eqn\redmod{\fhg\ =\ {\epsilon\over2\sqrt{\zeta}}\ -\ \tsconf(\theta)\ -\
 \tscsb(\theta,\zeta)}
where $\tscsb=(g^2/8\pi)\scsb$ and
\eqn\tsconfdef{\tsconf(\theta)\ =\ {u^4+1\over(u^2+1)^2}\ -\ {2u^4\log u^4
\over(u^2-1)(u^2+1)^3}\ ,\quad u^2=1+{1\over2\theta^2}+
\sqrt{{1\over4\theta^4}+{1\over2\theta^2}}\ .}
The saddle-point equations are then
\eqn\saddletheta{\ 0\ =\ {\partial\over\partial\theta}\tsconf\ -\
{\partial\over\partial\theta}\tscsb\ ,}
\eqn\saddlezeta{0\ =\ {\epsilon\over4\zeta^{3/2}}\ +\
{\partial\over\partial\zeta}\tscsb\ .}

The numerical evaluation of $\tscsb$ and its derivatives is actually
somewhat subtle, and we digress for a paragraph to discuss it. One
first performs the sum indicated in Eqs.~\scsbdef\ and \fnchoice\ in
closed form.  The angular part of the $dz\,d\bar z$ integration is
carried out analytically using contour methods. The subtle point is
that when $\theta$ is small, the resulting radial integral features a
sharp peak or boundary layer, whose contribution almost exactly
cancels that of the rest of the integration domain. This near
cancellation is carried out numerically to great accuracy with the
help of an appropriate rescaling of the boundary layer.

This having been done, we proceed first to Eq.~\saddletheta, as it is
independent of energy. Figure 1 shows the numerical result of this
equation for two different values of $\lambda/g^2,$ namely $.2$ and 2.
The intercept value $\zeta=1/2$ in the far-separation (and low-energy,
see Fig.~2) limit $\theta\rightarrow0$ is no surprise: in this limit
the Khoze-Ringwald model \newfhg\ becomes a good approximant for
\ourmodel, and also $\tsconf\approx1-2\theta^2,$ so that the resulting
saddle-point algebra is elementary.

\def\ecrit{\epsilon_{\rm crit}^{}}
Using Fig.~1 to eliminate $\theta$ in favor of $\zeta,$ we next solve
Eq.~\saddlezeta\ to obtain the saddle-point value of $\zeta$ as a
function of energy $\epsilon.$ The numerical results are plotted in
Fig.~2, again for $\lambda/g^2=.2$ and 2. In either case a bifurcation
is evident: beyond a critical energy $\ecrit(\lambda/g^2)$ there is no
solution for $\zeta.$ More accurately, for $\epsilon>\ecrit$ the
saddle-point value of $\zeta$ leaves the real axis as a complex
conjugate pair.

Finally, Fig.~3 reassembles the complete holy grail function $\fhg$, from
Eq.~\redmod, as a function of energy, restricted to the range $0\le\epsilon\le
\ecrit(\lambda/g^2).$ By design,
the two values of $\lambda/g^2$ we have chosen give very different
results. In the ``light-Higgs'' case $\lambda/g^2=.2,$ $\fhg$ rises to
zero, and so $\sanom$ loses its exponential suppression.  This case is
in qualitative agreement with the results of Ref.~\Dorey\ in the
single-instanton sector of the $\sigma$ model, as well as with
Ref.~\KRtwo\ in Electroweak theory. However, remembering our first
order perturbation theory criterion \fopt, we calculate that where
$\fhg\approx0$ the two terms $\scsb[w_V]$ and $\sconf[w_V]$ give
approximately equal contributions to $\fhg,$ strongly violating the
criterion \fopt. Therefore, beyond low energies, we have no reason to
trust the ``light Higgs'' result shown in Fig.~3, especially not in
the interesting regime where $\fhg$ nears zero. For still smaller
values of $\lambda/g^2$, $\fhg$ rises even faster with energy,
becoming considerably greater than zero, but the inequality \fopt\ is
even more badly violated. Our calculation in the ``light Higgs''
regime is not self-consistent, and deserves no further discussion.

On the other hand, for the ``heavy Higgs'' case $\lambda/g^2=2,$
$\fhg$ only rises around 15\% prior to $\ecrit,$ so that $\sanom$
remains exponentially suppressed. And in contrast to the ``light
Higgs'' case, here $\scsb[w_V]$ never exceeds around 15\% of
$\sconf[w_V]$ in this energy range, so that the criterion \fopt\ is
reasonably well obeyed. For still larger values of $\lambda/g^2$ the
trend continues: $\sanom$ loses progressively less of its exponential
suppression before bifurcating, while first order perturbation theory
becomes increasingly reliable.

In sum, we have exhibited a self-consistent one-parameter family of
models, parame- trized by large values of $\lambda/g^2$ (say, greater
than 2), which bifurcate at energies where $\sanom$ is still
exponentially suppressed, and for which first-order perturbation
theory appears to be a reasonable approximation.  Needless to say, a
parallel calculation in Electroweak theory, although quite intricate,
would be of great interest. In such a calculation, $\scsb$ would not
involve an arcane construction such as Eq.~\scsbdef, but would be
given by the Standard Model Higgs Lagrangian.

We conclude with three brief comments about the limitations of our
model, and of our understanding:

($i$) \rm How does one properly extrapolate $\sanom$ beyond the
bifurcation energy? We have no idea. The valley method has been
shown\refmark{\peace,\BalBraun} to provide the analytic continuation
in the variable $\rho^2/R^2$ of the so-called ``$R$-term method'' of
Khlebnikov, Rubakov and Tinyakov.\refmark\KRT This equivalence extends
the optical theorem to the case of $\ii$ backgrounds.  But at
$\epsilon=\ecrit$ the valley method ceases to be analytic, and
consequently, we have no good reason to believe that it has anything
to do with the total anomalous cross section $\sanom.$ A \it
conservative \rm guess would be that $\epsilon =\ecrit$ marks the end
of the exponential rise of $\sanom.$ In any event, new physics for
$\epsilon\ge\ecrit$ is obviously required.

($ii$) \rm What about multi-instanton configurations? These, also, can
be thought of as a bifurcation in the $\ii$ valley. While in the
present scenario for $\epsilon\ge\ecrit$ the number of collective
coordinate degrees of freedom jumps discontinuously ($\rho$ and $R$
become complex), so too in the multi-instanton scenario of
Zakharov\refmark\Zakharov and Maggiore and Shifman\refmark\Maggiore
the number of relevant degrees of freedom increases at some critical
energy $\epsilon^{}_{\rm ZMS}$ to encompass the collective coordinates
of long chains of alternating $I$'s and $\bar I$'s.  Presumably, the
important bifurcation is the one that happens first.  However, a
calculation of $\epsilon^{}_{\rm ZMS}$ in the present version
\ourmodel\ of the $\sigma$ model is beyond the scope of this Letter
($cf.$ Ref.~\Dorey).  We also note the possibility that a bifurcation
of the type discussed herein occurs separately in each multi-instanton
sector, and prior to the point where the $I\bar II\bar I$ contribution
(for example) catches up to the $I\bar I$ result. If that is the case,
then multi-instanton contributions can be safely ignored.

($iii$) \rm What about initial-state corrections? Recently much
attention has focused on the semiclassical description of
initial-state corrections in the Ringwald problem.\refmark\initial
These corrections are absent in the valley method, except to the
extent that the division between final-state and initial-state effects
is itself somewhat ambiguous.\refmark{\KT,\peace} However, if the
final-state valley corrections by themselves are understood to cut off
the rise of $\sanom$ at an exponentially suppressed value, through a
bifurcation or otherwise, then it is difficult for us to imagine that
the additional effect of the overlap of the hard initial state with
the valley could
\it enhance \rm $\sanom$ and render it observable.

We thank Nick Dorey for valuable input at all stages of this work.
\listrefs
\centerline{\bf Figure Captions}
\noindent 1. Energy-independent relation between the saddle-point
values of collective coordinates
$\theta$ and $\zeta,$ from Eq.~\saddletheta. Here and in the subsequent
figures, the solid line denotes the ``heavy Higgs'' case
$\lambda/g^2=2,$ while the broken line denotes the ``light Higgs''
case $\lambda/g^2=.2.$

\noindent 2. Saddle-point value of $\zeta$ as a function of
energy, from Eq.~\saddlezeta. The open circles at the terminus
of the curves mark the bifurcation points, $\epsilon=\ecrit.$

\noindent 3. The holy grail function $\fhg,$ from Eq.~\redmod,
for energies $\epsilon\le\ecrit.$
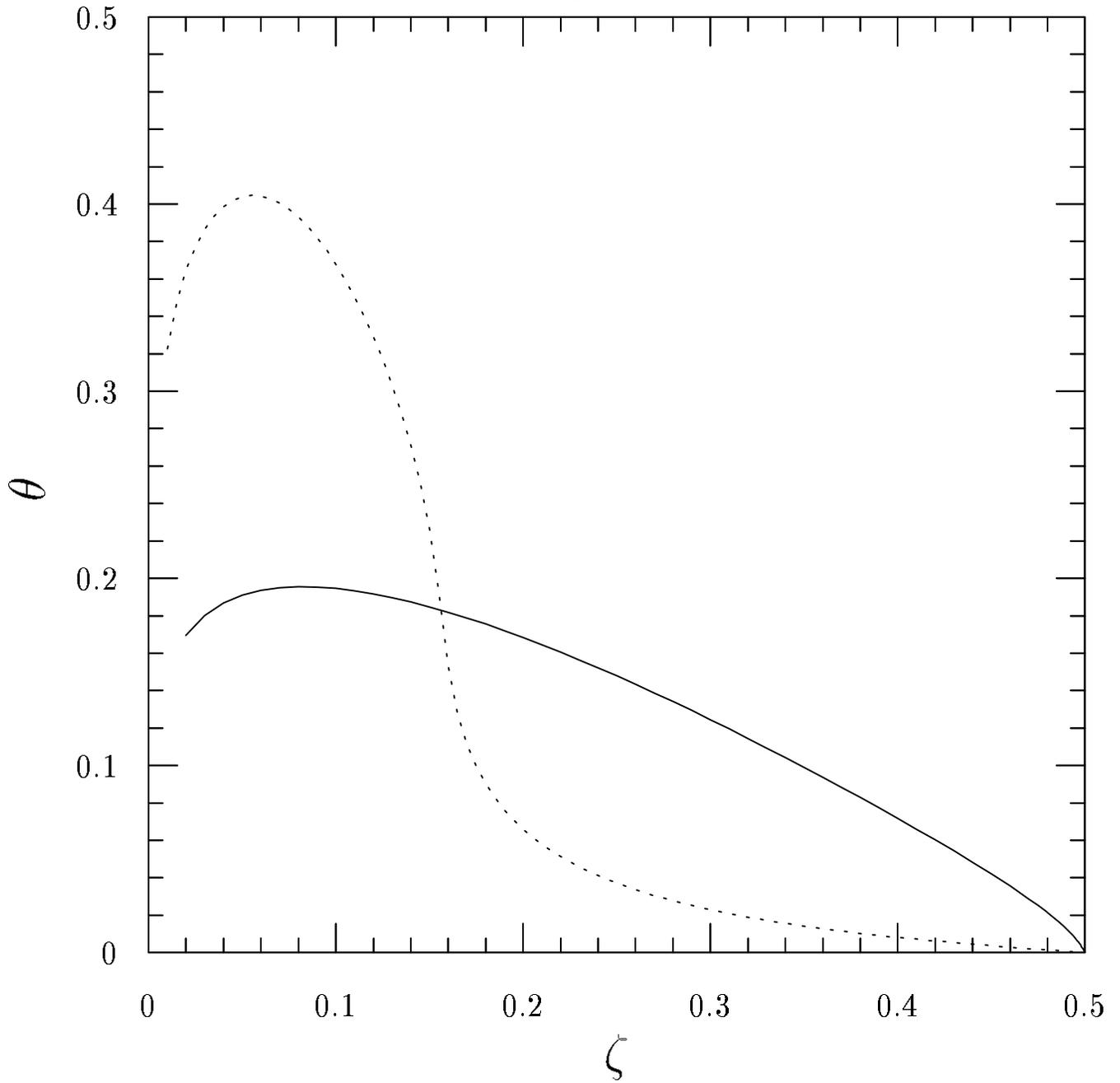
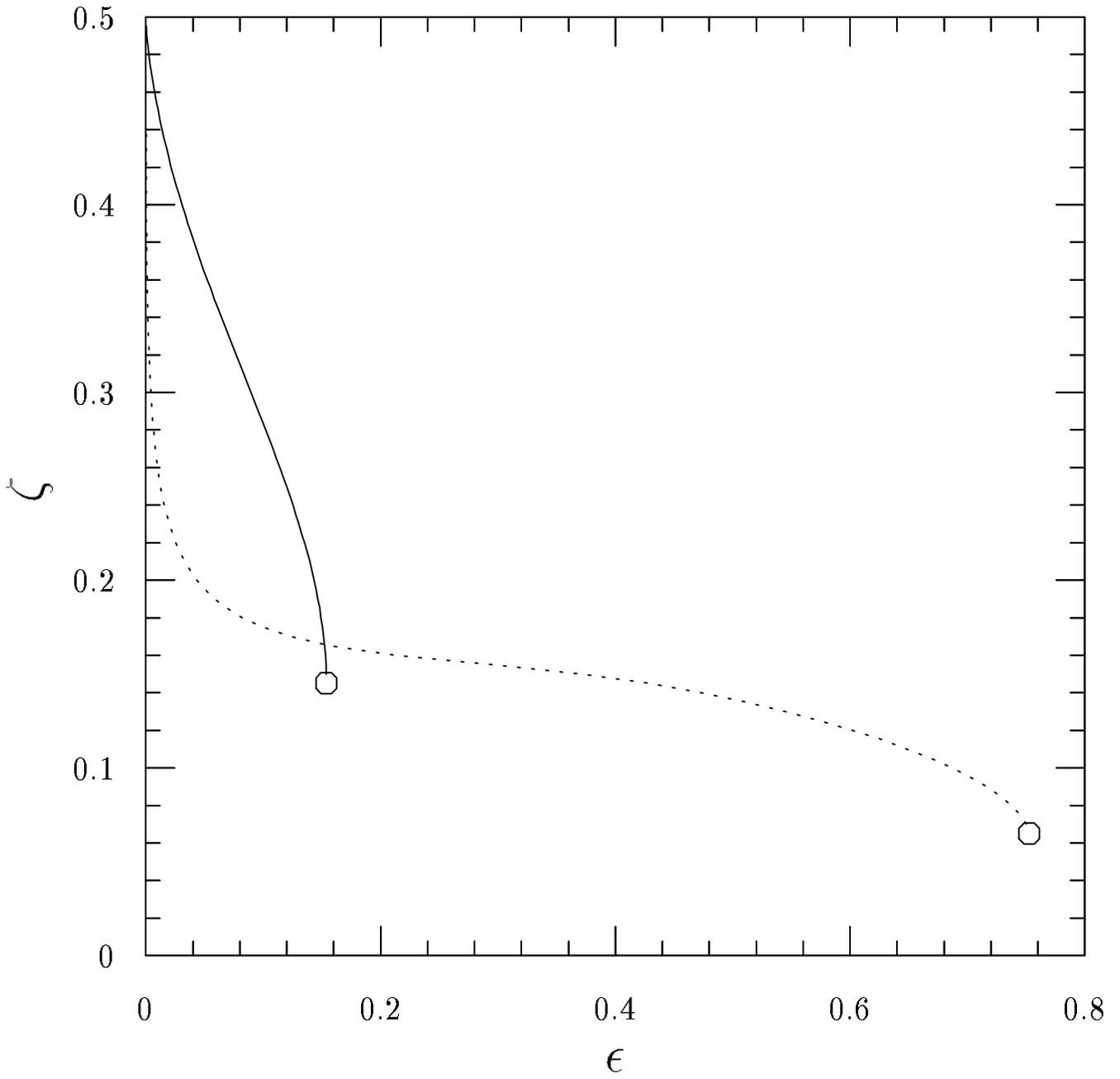
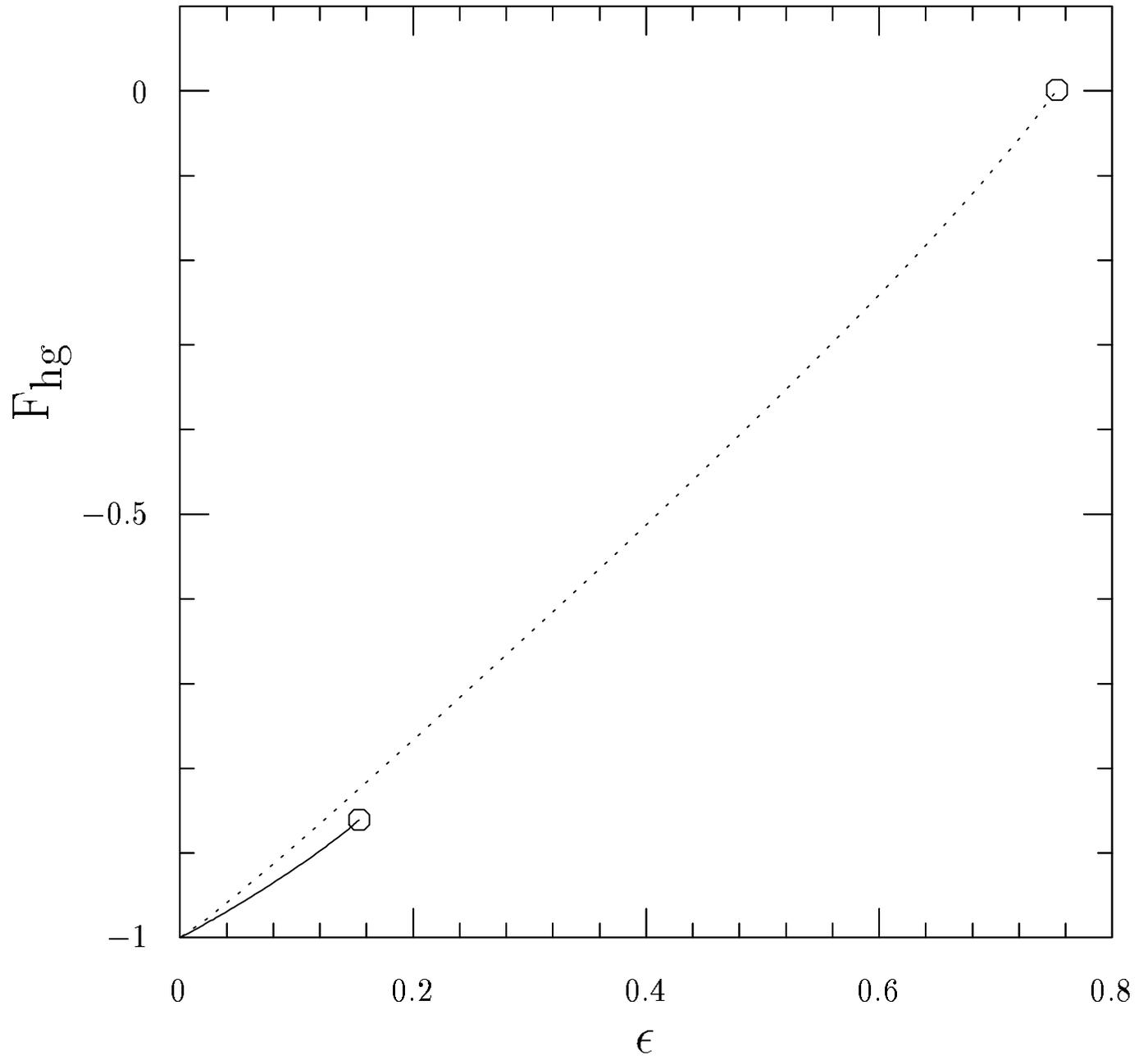
\bye